\documentclass[10pt,twocolumn,preprintnumbers,aps,amsmath,amssymb,prd,nofootinbib
,superscriptaddress]{revtex4-1}
\usepackage{graphicx,longtable,mathrsfs,color,array}
\usepackage[hidelinks]{hyperref}
\usepackage[usenames,dvipsnames]{xcolor} 
\usepackage{amssymb,amsmath,mathtools,mathrsfs,slashed} 
\usepackage{epsfig,subfigure,float} 
\usepackage{booktabs,longtable,ctable,multirow} 
\usepackage{exscale,relsize} 
\usepackage[normalem]{ulem} 
\usepackage{enumerate}
\usepackage{times, mathptmx} 
\usepackage[utf8]{inputenc}
\usepackage{color}
\usepackage{hyperref}
\usepackage{graphicx}
\usepackage{color}
\usepackage{physics}
\usepackage{graphicx,graphics}
\usepackage{bm}
\usepackage{tabularx}
\allowdisplaybreaks[1]
\usepackage[section]{placeins}

\newcommand{\A}{\delta A}
\newcommand{\ef}{v} 
\newcommand{\mtf}{{\varphi_m}} 
\newcommand{\scf}{\varphi} 
\newcommand{\rhoEF}{\rho_{\textup{EF}}} 
\newcommand{\Src}{\mathfrak{S}}

\begin{document}

\title{The quasi normal modes of growing dirty black holes}
\author{Jamie Bamber}
\email{james.bamber@physics.ox.ac.uk}
\affiliation{Astrophysics, University of Oxford, DWB, Keble Road, Oxford OX1 3RH, UK}
\author{Oliver J. Tattersall}
\email{oliver.tattersall@physics.ox.ac.uk}
\affiliation{Astrophysics, University of Oxford, DWB, Keble Road, Oxford OX1 3RH, UK}
\author{Katy Clough}
\email{katy.clough@physics.ox.ac.uk}
\affiliation{Astrophysics, University of Oxford, DWB, Keble Road, Oxford OX1 3RH, UK}
\author{Pedro G. Ferreira}
\email{pedro.ferreira@physics.ox.ac.uk}
\affiliation{Astrophysics, University of Oxford, DWB, Keble Road, Oxford OX1 3RH, UK}

\date{Received \today; published -- 00, 0000}

\begin{abstract}
The ringdown of a perturbed black hole contains fundamental information about space-time in the form of Quasi Normal Modes (QNM). Modifications to general relativity, or extended profiles of other fields surrounding the black hole, so called ``black hole hair", can perturb the QNM frequencies. Previous works have examined the QNM frequencies of spherically symmetric ``dirty" black holes -- that is black holes surrounded by arbitrary matter fields. Such analyses were restricted to static systems, making the assumption that the metric perturbation was independent of time. However, in most physical cases such black holes will actually be growing dynamically due to accretion of the surrounding matter. Here we develop a perturbative analytic method that allows us to compute for the first time the time dependent QNM deviations of such \textit{growing} dirty black holes. 
Whilst both are small, we show that the change in QNM frequency due to the accretion can be of the same order or larger than the change due to the static matter distribution itself, and therefore should not be neglected in such calculations.
We present the case of spherically symmetric accretion of a complex scalar field as an illustrative example, but the method has the potential to be extended to more complicated cases.

\end{abstract}

\maketitle


\section{Introduction}

The final stage of black hole formation, either from a binary merger or gravitational collapse, is a perturbed single black hole (BH) which ``rings" like a bell. The gravitational waves emitted during this ``ringdown" phase are dominated by a discrete set of damped oscillatory modes dubbed {\it Quasi Normal Modes} (QNM), whose frequencies are strictly determined by the underlying spacetime, and are indexed by overtone number $n$ and angular numbers $l,m$. In the case of standard general relativity (GR) and an isolated Kerr BH, the QNM frequencies are uniquely determined by the BH mass and spin. The detection of gravitational waves from binary mergers by the Advanced LIGO/Virgo network (recently augmented by the addition of KAGRA) \cite{Advanced_LIGO, VIRGO, Somiya:2011np}) provides a means by which to directly measure these QNM frequencies \cite{Aasi:2013wya, Carullo_2019,Giesler_2019,Isi_2019} and thus probe the spacetime around black holes directly. 
The prospects for this field of ``Black Hole Spectroscopy" will only improve as future detectors such as LISA and the Einstein Telescope come online \cite{LISA_BH_spec, Punturo:2010zz, Cabero_2020, Dreyer:2003bv, Berti:2015itd}. 

Methods for calculating and studying the QNMs of Kerr BHs in standard GR, both numerical and analytic, are well established \cite{Berti_2009,Konoplya_2011, Nollert:1999ji,Kokkotas:1999bd,Cho:2012, Ferrari:2007dd, Govindarajan:2000vq}, 
but only a few works have extended these techniques to cases of modified gravity or non trivial matter environments; so called ``dirty" or ``hairy" black holes \cite{QNM_dirty_1999, DirtyBH_QNM_2004, Nagar_2007, Barausse:2014tra, Nielsen:2019ekf, Matyjasek:2020bzc}.
A change in the black hole metric $\delta g_{\mu\nu}$, arising from modifications to GR or from the backreaction of surrounding matter, will result in a corresponding shift in QNM frequencies $\delta \omega_{nlm}$. Such effects are likely to be small \cite{Barausse:2014tra}, but have yet to be fully quantified. 

One simple and physically motivated situation in which there is non-zero hair around a black hole is where the BH accretes matter from the surrounding environment. Observational evidence of an electromagnetic counterpart to the gravitational wave event GW190521 suggested that it may be a binary black hole merger occuring within the accretion disc of an active galactic nucleus \cite{Graham:2020gwr}, meaning that ``dirty" black hole mergers in matter-rich environments are not an entirely theoretical concept. 

While baryonic accretion discs are perhaps the most well motivated example of matter accretion, in this paper we first examine the more straightforward case of spherically symmetric accretion. We use as an illustrative example the accretion of a massive complex scalar field onto a Schwarzschild BH, for which stationary solutions are known. Such an environment could describe a black hole located inside a bosonic dark matter halo \cite{Hui_2019, Clough:2019jpm, Bamber:2020bpu, Annulli:2020lyc, Annulli:2020ilw} or the end point of boson star mergers or collapses \cite{Palenzuela:2006wp, Palenzuela:2007dm, Bezares:2017mzk, Helfer:2016ljl, Helfer:2018vtq, Clough:2018exo, Widdicombe:2019woy, Widdicombe:2018oeo, Sanchis-Gual:2020mzb}, among other scenarios. Whilst such an accreting black hole is ultimately not a truly stationary state (at some point one would expect the asymptotic source of matter feeding the accretion to be ``used up''), over any short period of time the configuration is well-described by a steady state profile, with a fixed rate of flow into the horizon.

Even restricting to the case of spherical symmetry, calculating the QNM perturbations for such a growing, ``dirty" BH presents several novel challenges. The first is that since the matter is continually accreting onto the BH, the BH mass increases with time, and the metric deviation acquires a time dependence, $\delta g_{\mu\nu} = \delta g_{\mu\nu}(t,r)$. Most previous works have been limited to static metric shifts $\delta g_{\mu\nu}(r)$. Numerical results have been obtained for the quasi normal modes of scalar and electromagnetic perturbations in a time dependent Vaidya metric \cite{Abdalla:2006vb,Shao_2005,He:2009jd,Lin:2019fte}, however as far as we are aware perturbative analytic results for gravitational quasi normal modes on a time dependent background have not been obtained. 

The second challenge is that in the standard coordinate choice - Schwarzschild coordinates - the accreting matter piles up around the horizon because the time coordinate there is singular. A different choice is required to avoid the resulting divergence in the backreaction.

To overcome these challenges we combine and extend techniques from two previous works. Firstly, Cardoso et al. \cite{Cardoso_2019}, who demonstrated a procedure of re-definitions to produce modified QNM equations (i.e. modified Zerilli and Regge-Wheeler equations) for spherically symmetric and static metric shifts on a Schwarzschild background, and from this a numerical code for computing the QNM shifts. Secondly, Dolan \& Ottewill \cite{Dolan_2009}, who described a perturbative analytic technique for computing quasinormal modes of known static, spherically symmetric spacetimes. We combine these approaches to produce a novel way of computing QNMs, and show how the use of an adapted coordinate system can be used to tackle the accreting case.

This paper is organised as follows. In Sec. \ref{sec:backreaction} we set up the background spacetime of an accreting black hole. In Sec. \ref{sec:perturbed_QNM_eq} we derive the quasinormal mode equations for the perturbed metric. In Sec. \ref{sec:method}  we compute an analytic perturbative expression for the QNM deviations of a growing, dirty, Schwarzschild BH and in Sec. \ref{sec-accretingexamples} we give explicit results for massive complex scalar field ``dirt".  We conclude in Sec. \ref{sec:future_work}, discuss our results and propose directions for future work.  In a series of appendices we discuss, in more depth, key steps in our work and, in particular, in App. \ref{sec:SchMethod}  and \ref{sec:simple_examples} we verify our method by applying it to simpler, well studied examples for which we have numerical results, to demonstrate that our analytic method gives good agreement.

Throughout this paper we assume the metric signature $(-,+,+,+)$ and geometric units $G = c = 1$. 

\section{The perturbed metric of accreting dark matter}
\label{sec:backreaction}

Consider a situation in which one has a sufficient reservoir of material far from the BH such that the system can reach an equilibrium where the loss of matter into the BH is balanced by the infall of matter from infinity, forming a long lived quasi-stationary cloud. This massive cloud will perturb the metric, and thus change the frequency of quasi normal modes.

We can write the metric as
\begin{equation}
    g_{\mu\nu} = g^{(0)}_{\mu\nu} + \delta g_{\mu\nu}.
\end{equation}
where $g^{(0)}_{\mu\nu}$ is the Schwarzschild background metric and $\delta g$ is the matter induced perturbation. Let $\mtf$, for now, represent general ``matter'' fields that source Einstein's equations. The zeroth order field solution $\mtf^{(0)}$ satisfies the equations of motion on the Schwarzschild background  
\begin{equation}
	\nabla^{(0)}_{\mu} T^{\mu\nu}[\mtf^{(0)}, g^{(0)}_{ab}] = 0.
\end{equation}
The metric perturbation $\delta g_{\mu\nu}$ then satisfies 
\begin{equation}
	\delta G_{\mu\nu}[g^{(0)}_{ab},\delta g_{ab}] = 8 \pi T_{\mu\nu}[\mtf^{(0)}, g^{(0)}_{ab}], 
\end{equation}
where $\delta G_{\mu\nu}$ is the first order perturbation in the Einstein Tensor. 

For simplicity we will consider a spherically symmetric cloud on a spherically symmetric Schwarzschild background. Consider a diagonal perturbed line element of the form
\begin{equation}
	\dd s^2 = -(f + \delta f) \dd t^2 + (f + \delta g)^{-1} \dd r^2 + r^2 \dd \Omega, \label{eq:Sch}
\end{equation}
where $f(r) = 1 - 2M/r$, $\dd \Omega = \dd \theta^2 + \sin^2(\theta) \dd \phi^2$ and $M$ is the mass of the black hole. The perturbed Einstein field equations are then
\begin{align}
    \delta G^t_t &= \frac{1}{r^2}\partial_r(r\delta g) = 8\pi T^t_t,\label{eq:Gtt} \\ 
    \delta G^r_r &= (\delta g - \delta f)/(fr^2) + \frac{1}{r^2}\partial_r(r\delta f) = 8\pi T^r_r, \label{eq:Grr}\\
    \delta G^r_t &= -\partial_t \delta g/r = 8\pi T^r_t. \label{eq:Grt}
\end{align}

Now assume that the black hole is surrounded by a cloud of accreting matter described by a density $\rho := -T^t_t$. As the background Schwarzschild metric is static, conservation of energy implies that
\begin{equation}
    \partial_t (4\pi r^2 T^t_t) + \partial_r (4\pi r^2 T^r_t) = 0.
\end{equation}
(This can also be derived from Eqs. \eqref{eq:Gtt} \& \eqref{eq:Grt}). If the density is static $\rho = \rho(r)$ then $\partial_r(4\pi r^2 T^r_t) = 0$ hence $T^r_t = \A / (4\pi r^2)$ for some radially constant value $\A(t)$ which relates to the flux into the BH at some point in time $t$.

If we now choose to reparametrise $\delta g$ as $\delta g = - 2 \delta M(t,r)/r$, Eqs. \eqref{eq:Gtt} \& \eqref{eq:Grt} give
\begin{align}
    \partial_r \delta M &= 4 \pi r^2 \rho, \\
    \partial_t \delta M &= \A,
\end{align}
from which we can see that $\delta M$ is the additional effective mass of the black hole due to the cloud, and $\A$ is the rate of increase of mass of the BH due to accretion. Note that whilst in principle the quantities $\rho$ and $\A$ are independent, such that one could choose to have non zero density of matter near the horizon, but not have any flux into the BH, in most physical situations they will be related and of the same order. This can be seen explicitly in our illustrative example for a complex scalar field below, and explains our finding that the QNM frequency shift due to the accretion is of the same order as that due to the static matter distribution.

As alluded to in the introduction, the use of Schwarzschild coordinates presents problems for realistic examples. Consider our example case of a complex scalar field $\scf$ accreting onto a BH from an asymptotically constant energy density. 
As discussed in \cite{Hui_2019}, the stationary solution close to the horizon is
\begin{equation}
	\scf \rightarrow \scf_0 e^{-i\omega_s(t+r_*)} \quad r \rightarrow 2M
\end{equation}
and so
\begin{equation}
	\rho \rightarrow -\frac{2 \vert \scf_0 \vert^2 \omega_s^2}{1-2M/r} \quad r \rightarrow 2M
\end{equation}
diverges there. As a result our metric perturbation $\delta g$ also diverges, which breaks our assumption that $\delta g_{\mu\nu}$ is small. This is a typical result for matter distributions with a non-zero flux into the horizon, due to the coordinate singularity of the Schwarzschild metric at the horizon. 

The standard solution is to change to 
ingoing Eddington-Finkelstein (EF) coordinates, $\ef \equiv t + r_*$, where the tortoise coordinate $r_*$ is defined as 
\begin{align}
    \dd r_* =&~ \dd r / f(r), \\
    r_* =& ~r + 2 M \ln \left(\frac{r}{2M} - 1\right).
\end{align}
In the ingoing EF coordinates the Schwarzschild line element is 
\begin{equation}
    \dd s^2 = -f \dd \ef^2 + 2 \dd \ef \dd r + r^2 \dd \Omega ~.
\end{equation}
We define a perturbation in the metric $\delta \lambda (v,r)$ such that the line element is
\begin{equation}
    \dd s^2 = -F e^{2\delta \lambda(\ef,r)}\dd \ef^2 + 2e^{\delta \lambda(\ef,r)}\dd \ef \dd r + r^2 \dd \Omega,
\end{equation}
where 
\begin{equation}
	F = f - 2\delta M(\ef,r)/r,
\end{equation}
One can then show \cite{Babichev_2012} that similarly to the Schwarzschild case
\begin{align}
    \partial_r \delta M =& ~-4\pi r^2 T^\ef_{\ef} = 4\pi r^2 \rho_{EF}, \\
    \partial_{\ef} \delta M =& ~ -4\pi r^2 T^r_{\ef} = \delta A,
\end{align}
Where $\rhoEF = -T^\ef_{\ef}$ is the energy density measured by coordinate observers in ingoing EF coordinates, and $\delta A$ is the rate of increase in mass of the BH as before. 
In these coordinates the scalar field $\scf(\ef,r) \rightarrow \scf_0 e^{-i\omega_s \ef}$ as $r \rightarrow 2M$, so 
\begin{equation}
    \rhoEF = \left(f\vert \partial_r \scf \vert^2 + \mu^2 \vert \scf \vert^2\right) \rightarrow \mu^2 \vert \scf_0 \vert^2, \quad r \rightarrow 2M
\end{equation}
is perfectly well behaved at the horizon. We also have 
\begin{equation}
    \begin{split}
    \partial_r \delta \lambda =& -4\pi r T^\ef_r = - 4\pi r T_{rr} = \vert \partial_r \scf \vert^2, \\
        \rightarrow& 0, \quad r \rightarrow 2M.
\end{split}
\end{equation}
so the metric perturbation $\delta \lambda$ is also well behaved. For the scalar field we have explicitly
\begin{align}
    \delta A =& \; 8\pi (2M \omega_s)^2 \vert \scf_0 \vert^2, 
\end{align}
which, as expected, is independent of $r$ as the scalar field solution is stationary. In general, at any ($\ef,r$) we have
\begin{align}
	\delta M(\ef, r) =& \; \A ~ \ef + \int^r_{2M} 4\pi \bar{r}^2 \rhoEF ~ \dd \bar{r}, \label{eq:delta_M} \\
	\delta \lambda(\ef, r) =& - \int^r_{2M} 4\pi \bar{r} T_{rr} ~ \dd \bar{r}. \label{eq:lambda}
\end{align}
and for the specific case of the scalar field, we find
\begin{align}
	\delta M(v,r) =& \; 8\pi (2M \omega_s)^2 \vert \scf_0 \vert^2 v + \int^r_{2M} 4\pi \bar{r}^2 \left(f\vert \partial_{\bar{r}} \scf \vert^2 + \mu^2 \vert \scf \vert^2\right) \dd \bar{r}, \\
	\delta \lambda(r) =& -2 \int^r_{2M} 4\pi \bar{r} \vert \partial_{\bar{r}} \scf \vert^2 \dd \bar{r},
\end{align}
From this point on we will assume that, as in the case of the stationary scalar field solution, $\delta A$ is a constant and $\delta \lambda$ depends only on $r$.

In this section we have formulated the necessary expressions for the backreaction onto a Schwarzschild black hole due to stationary accretion in ingoing EF coordinates. We can now use the resulting metric to construct modified equations for the quasi normal modes.
Note that some further commentary and clarifications on the orders of perturbation required are provided in App. \ref{App-perturbationtheory}.

\section{Quasi normal mode equations on the perturbed background}
\label{sec:perturbed_QNM_eq}

A general spherically symmetric 4D spacetime can be written as the product of a 2D pseudo-Riemannian manifold $(\mathcal{M},\Tilde{g}_{ab})$ and the 2-sphere $(S_2,\hat{g}_{AB})$, 
\begin{equation}
    \dd s^2 = \Tilde{g}_{ab} ~\dd x^a \dd x^b + r^2 \hat{g}_{AB} ~\dd x^A \dd x^B,
\end{equation}
where indices $a,b \in \{\Tilde{t},r\}$, $A,B \in \{\theta,\phi\}$, with $\Tilde{t} = \Tilde{t}(t, r)$. 
It can be shown \cite{Brizuela_2008,Brodbeck_2000} that odd linear gravitational perturbations about such a metric can be described by a Regge-Wheeler-like master equation,
\begin{equation}
    \left[\Tilde{\nabla}_a \Tilde{\nabla}^a - V\right]\Psi = \Src, \label{Master_eq}
\end{equation}
where 
\begin{align}
    V =& \frac{(l+2)(l-1)}{r^2}+\frac{2}{r^2}\Tilde{g}^{rr}-\frac{1}{r}\Tilde{\nabla}_a \Tilde{\nabla}^a r, \\
    \Tilde{\nabla}_a \Tilde{\nabla}^a &= \frac{1}{\sqrt{-\Tilde{g}}}\partial_a\left(\sqrt{-\Tilde{g}}\Tilde{g}^{ab}\partial_b\right), 
\end{align}
and $\Src$ is a matter source term derived from $T_{\mu\nu}$. To find the quasi normal mode frequencies we solve the homogeneous equation with $\Src=0$
\footnote{Unfortunately an equivalent generalisation of the even mode Zerilli equation to non-vacuum backgrounds has not been found, so we will focus on the odd modes.}
. For the perturbed ingoing Eddington-Finkelstein metric $\Tilde{t} = v,~\Tilde{g}^{rr} = F, ~\sqrt{-\Tilde{g}}=e^{\delta \lambda}$, and so the homogeneous equation is 
\begin{align}
    \left[2 e^{-\delta \lambda} \partial_{\ef} \partial_r + e^{-\delta \lambda} \partial_r (e^{\delta \lambda}F \partial_r) - V\right]\Psi &= 0, \\
    \left[2 \partial_{\ef} \partial_r + \partial_r (F_*\partial_r) - V_*\right]\Psi &= 0, \label{eq:first_Psi}
\end{align}
where $F_* = e^{\delta \lambda} F$ and $\ef_* = e^{\delta \lambda} V$. Then
\begin{align}
    V_* =& e^{\delta \lambda}\frac{(l+2)(l-1)}{r^2}+\frac{2}{r^2}F_*-\frac{1}{r}F'_*, \\
    \approx& (1+\delta\lambda)V_- - \frac{6\delta M}{r^3}+\frac{2\delta M'}{r^2}-\delta \lambda' \frac{1}{r}f,
\end{align}
where the prime ${}'$ denotes $\partial_r$. If we take the effective BH horizon as being at $F_*(r) = 0$ then to first order the horizon radius is shifted from $2M$ to
\begin{equation}
\begin{split}
    r_H(\ef) &= 2M[1 + \delta M(\ef,2M)/M], \\
    &= 2M[1 + \A \ef/M].
\end{split}
\end{equation}
We can introduce a function $Z(\ef,r)$ such that
\begin{align}
    F_*(\ef,r) =& f_H(r)Z(\ef,r), \\
    f_H :=& 1 - r_H/r, \\
    Z \approx& 1 + \delta Z, \\
    \delta Z(r) \approx& - \frac{2(\delta M(\ef,r) - \delta M(\ef,2M))}{r-2M}+\delta \lambda(r),
\end{align}
where our choice of definition means that $\delta Z$ does not depend on $\ef$. Note that $\ef = 0$ is defined as when the effective horizon $r_H = 2M$. Following the method of \cite{Cardoso_2019}, we define $\Phi = \sqrt{Z}\Psi$. Then 
\begin{equation}
    \left[\partial_r (f_H \partial_r) + 2 (1-\delta Z)\partial_{\ef}\partial_r - \delta Z'\partial_{\ef} - \Tilde{V}\right]\Phi = 0,
    \label{eq:Phi_eq}
\end{equation}
where 
\begin{align}
    \Tilde{V} &= \frac{V_*}{Z} + \frac{1}{2}Z^{-\frac{1}{2}}(Z'Z^{-\frac{1}{2}}f_H)', \\
    \Tilde{V} &\approx V_* - \delta Z V_* + \tfrac{1}{2}(f \delta Z')', \\
    &\approx V_- + \frac{(l-2)(l+1)}{r^2}(\delta \lambda-\delta Z)+
    \tfrac{1}{2}(f \delta Z')'-\frac{f}{r} \delta Z',
\end{align}
to linear order in $\delta$. 

We now wish to solve for quasi normal mode solutions. For time independent metrics we look for solutions of the form $\Phi \sim e^{-i\omega t}u(r)$. We can write this in ingoing Eddington-Finkelstein coordinates as $\Phi \sim e^{-i\omega \ef}u(r)$, incorporating the factor of $e^{+i\omega r_*}$ into $u(r)$. However as the metric now has a small linear time dependence we need to allow for the frequency and the $u$ function to drift with $\ef$, 
\begin{equation}
    \Phi = \exp(-i\omega(\ef)\ef)u(r,\ef),
\end{equation}
where
\begin{equation}
    \omega(\ef) = \omega_0 + \delta \omega(\ef),
\end{equation}
and $\omega_0$ is the unperturbed Schwarzschild QNM frequency, giving
\begin{multline}
    \Big[\partial_r (f_H \partial_r) - 2i (\partial_{\ef}(\ef\omega) - \delta Z \omega_0) \partial_r \\
     - (\Tilde{V} - \delta Z' i\omega_0 ) + (2(1-\delta Z)\partial_r - \delta Z')\partial_{\ef} \Big] u(\ef,r) = 0.
\end{multline}
This expression cannot be directly solved using our method, which requires a differential equation in a single variable.
To enable this we introduce a ``comoving" coordinate $\Tilde{r}$ (from now on we use units where $M=1$), which we define as 
\begin{equation}
    \Tilde{r} = \frac{r}{(1+ \A \ef \sigma(r))}.
\end{equation}
where we choose function $\sigma$ such that $\sigma(r) \approx 1$ for $r \ll r_0$ and $\sigma \rightarrow 0$ for $r \rightarrow \infty$. The $r_0$ is some constant radius much larger than the black hole but much smaller than the distance between us and the black hole (one can think of it as the size of the accreting cloud). Then we can have for $r \ll r_0$
\begin{equation}
    \Tilde{r} \approx \frac{2r}{r_H}, \quad
    f_H \approx 1 - \frac{2}{\Tilde{r}},
\end{equation}
while for $r \gg r_0$ we recover $\Tilde{r} \approx r$. 
We now derive the QNM equation in terms of this single variable $\Tilde{r}$.
We have
\begin{align}
    \partial_{\ef} &= -\frac{\A r \sigma}{(1+ \A \ef \sigma)}\partial_{\Tilde{r}} + \partial_{\ef}, \\
    \partial_r &= \frac{1-\A\frac{\ef r \sigma'}{1+\A \ef \sigma}}{(1+ \A \ef \sigma)}\partial_{\Tilde{r}}.
\end{align}
where $\sigma' = \partial_r \sigma(r)$. Now let \begin{align}
    \kappa_r &:= \left(\pdv{\Tilde{r}}{r}\right)_{\ef} = \frac{1-\A\frac{\ef r \sigma'}{1+\A \ef \sigma}}{(1+ \A \ef \sigma)}, \\
    \kappa_{\ef} &:= \left(\pdv{\Tilde{r}}{v}\right)_r = -\frac{\A r \sigma}{(1+ \A \ef \sigma)}.
\end{align}

\begin{widetext}
If we now define $f_A = f_H + 2\frac{\kappa_{\ef}}{\kappa_r}(1-\delta Z)$, we have that
\begin{equation}
\begin{split}
    \Bigg[\partial_{\Tilde{r}} ( f_A \partial_{\Tilde{r}}) + 
    \kappa^{-1}_r\left(-2i(\partial_{\ef}(\ef\omega) - \delta Z(r) \omega_0) + \frac{\kappa_r'}{\kappa_r}f_A +\frac{\kappa_{\ef}}{\kappa_r}\delta Z'\right) \partial_{\Tilde{r}}  - \kappa^{-2}_r(\Tilde{V} - \delta Z'(r) i\omega_0 ) ~ + & \\
     \left(2(1-\delta Z(r))\kappa^{-1}_r \partial_{\Tilde{r}}-\kappa^{-2}_r\delta Z'(r)\right)\partial_{\ef} \Bigg]u(\ef,\Tilde{r}) &= 0. \label{eq:u}
\end{split}
\end{equation}
In the regime $r \ll r_0$ Eq. \eqref{eq:u} reduces to
\begin{equation}
\begin{split}
    \Big[\partial_{\Tilde{r}} (f_A \partial_{\Tilde{r}}) - r_H i (\partial_{\ef}(\ef\omega) - \delta Z(r) \omega_0) \partial_{\Tilde{r}} - 
     \frac{r^2_H}{4}(\Tilde{V} - \delta Z'(r) i\omega_0 ) +& \\
     \left(r_H(1-\delta Z(r))\partial_{\Tilde{r}}-\frac{r^2_H}{4} \delta Z'(r)\right)\partial_{\ef} \Big]u(\ef,\Tilde{r}) &= 0,
\end{split}
\end{equation}
\end{widetext}
to order $\delta^2$ where $f_A \approx 1 - 2/\Tilde{r} - 2\A \Tilde{r}$. We then have 
\begin{align}
    \delta Z(r) &\approx \delta Z(\Tilde{r}) + (r - \Tilde{r})\delta Z(r)' + \dots, \\
    &\approx \delta Z(\Tilde{r}) + \A \ef \sigma(r) r \delta Z(r)' + \dots, \\
    &\approx \delta Z(\Tilde{r}) + \mathcal{O}(\delta^2)
\end{align}
provided $\A \ef \sigma(r) r \ll 1$ for $r \ll r_0$. We can thus approximate $\delta Z(r) \approx \delta Z(\Tilde{r})$, and do the same for other order $\delta$ quantities. This means we can approximate
\begin{equation}
    \frac{r^2_H}{4}(\Tilde{V} - i\omega_0 \delta Z' ) \approx (\Tilde{V}(\Tilde{r}) - i \omega_0 \delta Z'(\Tilde{r})),
\end{equation}
where primes now denote $\partial_{\Tilde{r}}$. If we now look for solutions where $u = u(\Tilde{r})$ we have that 
\begin{equation}
    \Omega = \frac{r_H}{2}\partial_{\ef}(\ef\omega)-\omega_0 \delta Z(2), \label{eq:Omega_def}
\end{equation}
is independent of $\ef$. This gives us an equation in a single variable, 
\begin{multline}
    \Big[\partial_{\Tilde{r}} (f_A \partial_{\Tilde{r}}) - 2 i (\Omega - [\delta Z-\delta Z(2)] \omega_0)\partial_{\Tilde{r}} \\
    - (\Tilde{V}(\Tilde{r}) - i \omega_0 \delta Z'(\Tilde{r}))\Big]u(\Tilde{r}) = 0. 
\end{multline}
We can further simplify this by letting 
\begin{equation}
    u(\Tilde{r}) = \exp\left(-i\omega_0 \int^{\Tilde{r}} [\delta Z(\Tilde{r})-\delta Z(2)] / f_H(\Tilde{r}) ~\dd \Tilde{r} \right)\Tilde{u}(\Tilde{r}),
\end{equation}
(note that the exponent is well behaved as $\Tilde{r} \rightarrow 2$). Again neglecting $\mathcal{O}(\delta^2)$ terms this gives us
\begin{equation}
    \left[\partial_{\Tilde{r}} (f_A \partial_{\Tilde{r}}) - 2 i \Omega \partial_{\Tilde{r}} - (V_-(\Tilde{r})+\Delta V(\Tilde{r}))\right]\Tilde{u}(\Tilde{r}) = 0, \label{eq:final_u}
\end{equation}
where $\Delta V(\Tilde{r})$ contains all the potential terms of order $\delta$.

The aim of this section was to derive a differential equation in a single co-moving variable, for odd quasi normal modes about the perturbed Schwarzschild metric associated with the growing dirty black hole we described in the previous section. The equation derived, Eq. \eqref{eq:final_u}, can now be used to compute the quasi normal mode frequencies.

\section{Perturbative method for computing the quasi normal modes}
\label{sec:method}

While there are a host of numerical methods for calculating quasi normal mode spectra, here we adapt the method of \cite{Dolan_2009} to compute analytic, perturbative expressions for the corrections to the spectra described by Eq. \eqref{eq:final_u} for the fundamental $n=0$ modes. Due to the spherical symmetry the frequencies are independent of the spherical harmonic number, $m$. For $\Tilde{u}(\Tilde{r})$ we take an anzatz of the form
\begin{equation}
    \Tilde{u} = \exp\left(i\Omega \int^{\Tilde{r}_*} Y(\Tilde{r}) \dd \Tilde{r}_*\right) q(\Tilde{r}), \label{eq:u_anzatz}
\end{equation}
The principle idea is to expand in powers of $L = l + \tfrac{1}{2}$ so that
\begin{align}
    \Omega &= L\Omega_{-1} + \Omega + L^{-1}\Omega_1 + \dots, \\
    q(\Tilde{r}) &= \exp(S_0(\Tilde{r}) + L^{-1}S_1(\Tilde{r}) + \dots).
\end{align}
Substituting \eqref{eq:u_anzatz} into \eqref{eq:final_u}, we find the modified Regge-Wheeler equation takes the form
\begin{equation}
\begin{split}
    &f_A q'' + [f_A' + 2i\Omega(Y - 1)]q' + \\
    & \ \ \ \left[i\Omega Y' + \Omega^2 \frac{2Y-Y^2}{f_A} - \frac{L^2}{\Tilde{r}^2} - V_0 - \Delta V\right]q = 0, \label{eq:qY}
\end{split}
\end{equation}
where $V_0 = \frac{1}{\Tilde{r}^2}(-6/\Tilde{r} - 1/4)$ and 
\begin{equation}
\begin{split}
    \Delta V =& \frac{L^2-9/4}{\Tilde{r}^2}(\delta \lambda-\delta Z)+\tfrac{1}{2}(f_H\delta Z')'
    -\frac{f_H}{\Tilde{r}} \delta Z'\\
    &+\frac{2\omega_0^2}{f_H}[\delta Z(r)-\delta Z(2)].
\end{split}
\end{equation}
We can now match terms in orders of $L$. At order $L^2$ we have that
\begin{align}
     (2Y - Y^2)\Omega^2_{-1} =& \frac{f_A}{\Tilde{r}^2}, \\
    1 - (Y-1)^2 =& \frac{f_A}{\Omega^2_{-1} \Tilde{r}^2}, \\
    Y =& 1 \pm \left(1 - \frac{f_A}{\Omega^2_{-1} \Tilde{r}^2}\right)^{1/2}.
\end{align}

Let us now focus on the quasi normal mode boundary conditions. We want ingoing modes at the horizon and outgoing at $r \rightarrow \infty$, so
\begin{align}
    \Tilde{u} \sim& \;\Phi_0,  \quad & r_* \rightarrow -\infty, \\
    \Tilde{u} \sim& \; \Phi_{out} e^{2i\Omega r_*}, \quad &r_* \rightarrow \infty.
\end{align}
If we take $\delta A = 0$ and $\Omega_{-1} = 1/\sqrt{27}$ as in \cite{Dolan_2009} then $\Tilde{r} = r$ and
\begin{equation}
    \left(1 - \frac{f_A}{\Omega^2_{-1} \Tilde{r}^2}\right)^{1/2} = \left(1 - \frac{f 27}{r^2}\right)^{1/2} = \pm \left(1 - \frac{3}{r}\right)\left(1 + \frac{6}{r}\right)^{1/2}.
\end{equation}
We then can obtain the correct boundary conditions by taking 
\begin{equation}
    Y = 1 + \left(1 - \frac{3}{r}\right)\left(1 + \frac{6}{r}\right)^{1/2}.
\end{equation}
This is a modified form of the ansatz used in \cite{Dolan_2009} corrected for the fact that we have included a factor of $r_*$ into the $\ef$. 


We can try something similar with $\A \neq 0$. For small $r$ (i.e. $r \ll r_0$) we find 
\begin{equation}
    f_A \approx f_H -2\A \Tilde{r} = 1 - \frac{2}{\Tilde{r}} - 2 \A \Tilde{r}.
\end{equation}
If $\Omega_{-1} = 1/\sqrt{27} + \delta \Omega_{-1}$, one can show that
\begin{align}
    Y =& 1 + \left(1-\frac{3(1+3\A)}{\Tilde{r}}\right)\left(1+\frac{6(1+12\A)}{\Tilde{r}}\right)^{1/2}, \\
    \delta \Omega_{-1} =& -\sqrt{3}\A,
\end{align}
satisfies the order $L^2$ equation to order $\delta$. The repeated root for $Y = 1$ corresponds to the null unstable circular orbit, which shifts to $r = 3\frac{r_H}{2M}(1+3\A)$. We also note that $f_A$ and $Y(\Tilde{r})$ have zeros at $\Tilde{r} = 2+8\A+\mathcal{O}(\A^2)$ instead of at $\Tilde{r}=2$; however in the expression
\begin{equation}
    \int^{\Tilde{r}_*} Y(\Tilde{r})\dd \Tilde{r}_* = \int^{\Tilde{r}} Y(\Tilde{r})/f_A(\Tilde{r}) ~\dd \Tilde{r}
\end{equation}
these zeros cancel so that the integral is well behaved. Hence for well behaved $q(\Tilde{r})$ we obtain the correct boundary condition at $r \rightarrow r_H$. 

Now let us examine the limit of large $r$ (i.e. $r \gg r_0$). We want to confirm we obtain outgoing waves, i.e. 
\begin{equation}
    \Tilde{u}(\Tilde{r}) \approx \exp(2i\Omega r_*)
\end{equation}
as $r \rightarrow \infty$. We have $Y \approx 2 - f_A/(2\Omega^2_{-1}\Tilde{r}^2)+\dots$ so 
\begin{equation}
    \Tilde{u} \rightarrow \exp\left(i\Omega[ 2\Tilde{r}_*+\frac{1}{\Omega_{-1}\Tilde{r}}+\dots]\right)q(\Tilde{r})
\end{equation}
and 
\begin{align}
    \Tilde{r}_* &\approx \int \left(1 - \frac{2(1+\A \ef)}{r}+\A(\ef(\sigma r)'-2\sigma r) + \mathcal{O}(\A^2)\right)^{-1} \dd r, \\
    &\approx r + 2\ln(r-2) + \A\left(\ef(2\ln(r)-(\sigma r)')-2\sigma r\right)+\mathcal{O}(\A^2), \\
    &\approx r_* + \mathcal{O}(\delta)
\end{align}
Thus our anzatz does give us outgoing waves at large $r$ provided $q(\Tilde{r})$ is suitably well behaved and $\sigma$ goes to zero with large $r$ suitably fast. 

Having established our modified anzatz still satisfies the quasi normal mode boundary conditions we can go back to solving for the $\Omega_n, S_n$ terms, using the $r \ll r_0$ limit. At order $L^1$ we have
\begin{equation}
   2iS_0'(Y - 1) + i Y' + \frac{2 \Omega_0}{\Omega^2_{-1}\Tilde{r}^2} = 0.
\end{equation}
If we require that $S_m$ be continuous and differentiable at the null unstable orbit, at $\Tilde{r}_c=3(1+3\A)$, then setting $\Tilde{r}=\Tilde{r}_c$ we find
\begin{equation}
    \Omega_0 = -i\frac{\sqrt{3}}{2}\Omega_{-1} Y'(\Tilde{r}_c) = -\frac{i}{2\sqrt{27}} +\frac{2i}{\sqrt{3}}\A + \mathcal{O}(\delta^2).
\end{equation}
We can also extract $S_0'$ as 
\begin{equation}
    S_0'(\Tilde{r}) = i \frac{iY'+2\Omega_0/(\Omega_{-1}\Tilde{r})^2}{2(Y-1)}.
\end{equation}
At order $L^0$ we have
\begin{equation}
\begin{split}
    f_A({S'_0}^2+S''_0)+(f_A'+i\omega_0\delta Z)S_0'+\frac{2\Omega_1\Omega_{-1}-\Omega^2_0}{\Tilde{r}^2\Omega^2_{-1}}\\+2iS_1'\Omega_{-1}(Y-1)-V_0-\Delta V = 0.
\end{split}
\end{equation}
We can again set $\Tilde{r} = \Tilde{r}_c$ to find $\Omega_1$, and then rearrange to obtain the function $S_1'$. 

The above procedure can be repeated to obtain higher order terms. For order $L^{-n}, n \geq 1$ the general expression is
\begin{multline}
    f_A(\sum^n_{m=0}S'_m S'_{n-m}+S''_n) \\
    +(f_A'+i\omega_0\delta Z)S_n'+\frac{1}{\Tilde{r}^2\Omega^2_{-1}}\sum_{m=-1}^{n+1}\Omega_m \Omega_{n-m}\\
    +i\Omega_n Y'+2i(Y-1)\sum_{m=-1}^{n}\Omega_m S'_{n-m} = 0.
\end{multline}

The next few $\Omega_n$ terms are explicitly given by:
\begin{widetext}
\begin{align}
    \sqrt{27}\Omega_1 =& - \tfrac{281}{216}+\tfrac{1135}{72}\A+ \tfrac{9}{2} \Delta V(3), \\
    \sqrt{27}\Omega_2 =& i\tfrac{1591}{7776} - i\tfrac{1591
   }{432}\A + i\left(\tfrac{27}{8}\Delta V''(3)+9
   \Delta V'(3)+\tfrac{15}{4} \Delta V(3)\right), \\
   \sqrt{27}\Omega_3 =& -\tfrac{710185}{1259712}+\tfrac{2922805}{419904}\A-\tfrac{81}{64}
   \Delta V^{(4)}(3)-\tfrac{21}{2} \Delta V^{(3)}(3)-\tfrac{185
   }{8}\Delta V''(3)-\tfrac{29}{3}\Delta V'(3)+\tfrac{1061
   }{144}\Delta V(3), \\
   \sqrt{27}\Omega_4 =& i\tfrac{92347783
   }{362797056} - i\tfrac{69151003}{15116544}\A +i\Big[-\tfrac{81}{256}
   \Delta V^{(6)}(3)-\tfrac{171}{32}
   \Delta V^{(5)}(3)-\tfrac{1845}{64}
   \Delta V^{(4)}(3)-\tfrac{341}{6}
   \Delta V^{(3)}(3)-\tfrac{8087}{384}
   \Delta V''(3)+\tfrac{449}{16} \Delta V'(3) \nonumber \\
   &+\tfrac{16331
   }{1728}\Delta V(3)\Big], \\
   \sqrt{27}\Omega_5 =& -\tfrac{7827932509}{39182082048}-\tfrac{1376065091}{1451188224}\A +\tfrac{243
   }{4096}\Delta V^{(8)}(3)+\tfrac{27}{16}
   \Delta V^{(7)}(3)+\tfrac{8675}{512}
   \Delta V^{(6)}(3)+\tfrac{71519}{960}
   \Delta V^{(5)}(3)+\tfrac{1259827
   }{9216}\Delta V^{(4)}(3) \nonumber \\
   &+\tfrac{255217
   }{5184}\Delta V^{(3)}(3)-\tfrac{8562439
   }{93312}\Delta V''(3)-\tfrac{2427761
   }{69984}\Delta V'(3)+\tfrac{7696651
   }{419904}\Delta V(3), \\
   \dots \nonumber
\end{align}
\end{widetext}
Note that the terms zeroth order in $\delta$ are the same as  \cite{Dolan_2009}, showing we obtain the correct unperturbed frequency.

If we now examine Eq. \eqref{eq:Omega_def}, we have that
\begin{align}
    \partial_{\ef} (\ef\omega) &= (\Omega + \omega_0 \delta Z(2))/(1 + \A \ef),
\end{align}
which, when integrated and Taylor expanded gives us
\begin{align}
    \delta \omega &= \delta \Omega + \omega_0 \delta Z(2) - \omega_0 \tfrac{1}{2}\A \ef + \mathcal{O}(\delta^2).
\end{align}
We can rewrite this as
\begin{equation}
    \delta \omega = \delta \varpi + \delta A (\varpi_{A} - \omega_0 \tfrac{1}{2} \ef) + \mathcal{O}(\delta^2),
\end{equation}
where $\delta \varpi$ is the correction with zero accretion (arising from the static matter distribution around the horizon) and $\delta A(\varpi_A-\omega_0 \tfrac{1}{2}\ef)$ is the accretion term. 

So far we have obtained a solution in our adapted EF coordinates of the form
\begin{equation}
    \Phi = \exp \left(-i\left[\omega_0 + \delta \varpi + \A (\varpi_{A} - \frac{\omega_0}{2}\ef/M) \right]\ef\right)u\left(\Tilde{r}\right).
\end{equation}
where we have restored the factors of $M$. However, if we were to measure this scalar we would of course do so as asymptotic observers in $t$ and $r$ coordinates not $\ef$ and $\Tilde{r}$. Suppose we measure the scalar fluctuation at some fixed distance $r = R \gg 2M$ from the black hole, then $\ef = R_* + t$ only changes with $t$. Hence we can define a local time $t' = t - R_*$ such that at our position we just have $\ef = t'$. The origin at $t' = \ef = 0$ is defined by the point at which the BH mass is defined to be $M$, that is, when the effective black hole horizon $r_H = 2M$. We have also defined our $\Tilde{r}$ coordinate such that $\Tilde{r} \rightarrow r$ for $r \gg r_0$, so we observe
\begin{equation}
    \Phi \propto \exp \left(-i\left[\omega_0 + \delta \varpi + \A (\varpi_{A} - \frac{\omega_0}{2}t'/M) \right]t'\right)u(R).
\end{equation}
Note that as we only care about the time variation of the scalar at a fixed $R$ we can be agnostic about the precise form of $\sigma$, only assuming that it obeys the correct asymptotic boundary conditions.

Previous works have calculated $\delta \varpi$, assuming the accretion terms are zero. Yet as we shall see in the example of the massive complex scalar field in the next section, the contribution from the accretion terms $\A(\varpi_A - \frac{\omega_0}{2}t'/M)$ can in fact be larger than $\delta \varpi$. In the general case we would expect them to be at least of the same order, therefore the latter should not be neglected.

In this section we have arrived at our key result: a general formula for odd parity quasi normal mode style solution for growing dirty black holes, which can be used to study frequencies which are perturbed by the accretion of matter and how they drift with time. It is useful to check our approach, and in particular the expansion order required for accuracy, in simpler regimes where quasi normal mode frequencies have been calculated using other methods; we do so in App. \ref{sec:simple_examples}. We find that, in the static cases considered, the method is highly accurate to the 5th order expansion used, and that going to higher orders does not result in significant corrections. In the following section we apply the result to our illustrative example of scalar field accretion.

\section{Complex massive scalar field accretion} 
\label{sec-accretingexamples}

Having set up the framework and formalism, we can now apply it to our test case: a complex massive scalar field. Once again we set $M = 1$. From \cite{Hui_2019}  we can approximate the $\scf$ solution for small $M\mu < 1$ and $\omega_s=\mu$ as 
\begin{align}
    \scf \sim& \; \scf_0 e^{-i \mu(\ef-r)}, \quad 2 < r \lesssim \mu^{-2}/2, \\
    \scf \propto& \; r^{-3/4} e^{-i\mu(\ef-r_*)}\cos(2\mu\sqrt{2r} - 3\pi/4), ~ \mu^{-2}/2 \lesssim r.
\end{align}
In the regime where $r < \mu^{-2}/2$, from Eqs. \eqref{eq:delta_M} and \eqref{eq:lambda} we have
\begin{align}
    \delta A \approx& \;32 \pi \mu^2 \vert \scf_0 \vert^2 = 32 \pi \rho_h, \\
    \delta M \approx& \;32 \pi \rho_h\left[ \ef + \frac{1}{24}\left(2 r^3 - 3 r^2 - 4\right)\right], \\
    \delta \lambda \approx& \; 4 \pi \rho_h \left[4-r^2\right], \\
    \delta Z \approx& \; \frac{4}{3}\pi\rho_h\left[-7r^2-2 r+8\right],
\end{align}
which in turn gives
\begin{equation}
\begin{split}
    \Delta V(\Tilde{r}) = 8 \pi \rho_h \Big[&\frac{L^2}{3}\left(2+\Tilde{r}^{-1}+2\Tilde{r}^{-2}\right)-\frac{1}{6}(7\Tilde{r}^2 + 16\Tilde{r})\omega^2_0 \\
    &-\frac{61}{12}\Tilde{r}^{-1}-\frac{5}{2}\Tilde{r}^{-2}-\frac{1}{3}\Big].
\end{split}
\end{equation}
where we have expressed all quantities in terms of $\rho_h$, the scalar field density on the horizon in ingoing EF coordinates. 
Substituting in these values of $\Delta V$ and $\delta Z$ gives the perturbations to the quasi normal mode frequency as
\begin{equation}
\begin{split}
    &\delta \varpi = \frac{32 \pi \rho_h}{M^3}
   \Big[0.0360844 L+ 0.0160375 i\\
   &-0.0522147 L^{-1}-0.0222155 i L^{-2}-0.105189 L^{-3}+0.0307956
   i L^{-4}\\
   &-0.245579 L^{-5}+\mathcal{O}(L^{-6})\Big]
\end{split}
\end{equation}
and
\begin{equation}
\begin{split}
    &\A \varpi_A = \frac{32 \pi \rho_h}{M^3}
   \Big[-1.73205 L+ 1.1547 i\\
   &+3.03376 L^{-1}-0.708769 i L^{-2}+1.33958 L^{-3}-0.880368
   i L^{-4}\\
   &-0.182488 L^{-5}+\mathcal{O}(L^{-6})\Big]
\end{split}
\end{equation}
where we have now restored the factors of $M$. We note that the accretion term, $\A \varpi_A$, is in fact substantially larger than the non-accretion term, $\delta \varpi$, showing the importance of properly accounting for the accretion and the time dependence of the backreaction.

Putting both contributions together we obtain
\begin{equation}
\begin{split}
    &\delta \varpi + \A \varpi_{A} = \frac{32 \pi \rho_h}{M^3}
   \Big[-1.69597 L+1.13866 i\\
   &+2.98155 L^{-1}-0.730984 i L^{-2}+1.23439 L^{-3}-0.849572
   i L^{-4}\\
   &-0.428067 L^{-5}+\mathcal{O}(L^{-6})\Big]
\end{split}
\end{equation}
For comparison the equivalent expression for $\omega_0$ is
\begin{equation}
\begin{split}
    \omega_0 = M{^{-1}}\Big[&0.19245 L - 0.096225 i - 0.250363 L^{-1} \\
    &+ 0.039376 i L^{-2} - 0.108497 L^{-3} + 0.048987 i L^{-4} \\
    &- 0.0384483 L^{-5} +\mathcal{O}(L^{-6})\Big].
\end{split}
\end{equation}
Suppose we can measure the QNM ringdown signal for $N$ oscillations before it passes below our sensitivity threshold. Then the shift in frequency with time over the course of the ringdown will be order 
\begin{equation}
    - \delta A \frac{\omega_0}{2} \frac{\Delta t'}{M} \sim - \pi N \frac{\delta A}{M} = \pi N \left(\frac{32 \pi \rho_h}{M^3}\right). 
\end{equation}
Hence if $N$ is order 1, for a detector with a decent signal to noise ratio, then this shift with time is of a comparable size to the constant frequency change $\delta \varpi + \A \varpi_A$.

We can now assess how large this shift in QNM frequency actually is. For the complex scalar field the size of the deviations can be parameterised by the non-zero dimensionless accretion rate $\A$ which is related to the density on the horizon as described above
\begin{equation}
    \frac{\delta \omega}{\omega_0} \sim \frac{\delta \varpi + \A \varpi_A - \tfrac{1}{2}\delta A \omega_0 t'/M}{\omega_0} \sim \A = \left(\frac{32 \pi \rho_h}{M^2}\right).
\end{equation}

Plugging in the fundamental constants we find that a fractional BH mass growth rate of $ 10^{-14}\textup{Gyr}^{-1}$ corresponds to a $\delta A$ of $(M/M_{\odot})^2 10^{-6}$. Assuming that the complex scalar is dark matter, we can also express $\delta A$ in terms of a typical asymptotic dark matter density, as follows.
For the massive scalar field with $\omega = \mu$ the density decays as $\sim r^{-3/2}$ at large $r$ so we find
\begin{equation}
    \rho_h \sim \rho_{R_c} \pi (\mu M)^3 \left(\frac{2 R_c}{M}\right)^{3/2},
\end{equation}
where $\rho_{R_c}$ is the density at some large radius $R_c$ which we take to be the effective radius of the cloud. If we set $R_c$ by equating the virial velocity to a typical dispersion velocity of dark matter 
\begin{equation}
    R_c/M \sim (v_{\textup{disp}})^{-2} \sim \left(\frac{100 \textup{km s}^{-1}}{c}\right)^{-2} \sim 10^6, 
\end{equation}
then
\begin{equation}
    \A \sim 10^{-30}(\mu M)^3 \left(\frac{\rho_{R_c}}{M_{\odot}\textup{pc}^{-3}}\right).
\end{equation}
Hence we see that for an asymptotic scalar field mass density of $\sim 1 M_{\odot} \textup{pc}^{-3}$ and $M\mu < 1$ this is a very small effect. However in more extreme environments with larger matter densities or steeper density profiles the accretion may provide a more significant contribution to the frequency shift. 

Whilst we have chosen a specific form as an illustrative example, one can easily choose different profiles for $\scf$, or indeed different $T_{\mu\nu}$ profiles, compute $\Delta V$, and substitute into the expressions for $\Omega$ to get the corresponding QNM frequency shifts. 



\section{Discussion}
\label{sec:future_work}

While previous authors have attempted to estimate the QNM frequencies for ``dirty" black holes, that is black holes where the metric is perturbed by a stationary or quasi-stationary cloud of matter, their analyses were limited to simple, static, spherically symmetric metric perturbations around a Schwarzschild black hole \cite{QNM_dirty_1999,DirtyBH_QNM_2004,Barausse:2014tra}. 

However, in most physical cases such a cloud results in a steady flow of matter falling into the black hole, causing the mass of the black hole and the perturbed metric to acquire a time dependence. Here we present a perturbative analytic method to estimate, for the first time, the time dependent quasi normal mode frequencies for such a growing dirty black hole in spherical symmetry, assuming a linear time dependence. This method is based on the perturbative method of Dolan (2009) \cite{Dolan_2009} and the techniques for dealing with perturbed Schwarzschild metrics described in Cardoso et al. (2019) \cite{Cardoso_2019}. While the formula we derive can be applied to any kind of matter cloud, we give an illustrative result for a massive complex scalar field, in the context of wave-like dark matter. 
For our example we find that the size of the expected frequency shifts $\delta \omega$ can be related by the matter density near the BH horizon. We find that the frequency correction due to the time dependence of the metric, which other authors have neglected, is in fact \textit{larger} than the contribution from the static matter distribution. While these frequency shifts are tiny for typical astrophysical dark matter densities, it is possible that they could become relevant in very dense astrophysical environments.

Further details of the method are contained in the appendices. In particular, in App. \ref{sec:SchMethod} and \ref{sec:simple_examples} we have verified our method by applying it to several well studied static perturbed Schwarzschild space times, including generic potential deviations around a Schwarzschild background, a charged Reissner-Nordstr{\"o}m black hole and a Schwarzschild de Sitter black hole, and compared to previous numerical results where available. We find excellent agreement with previous results, demonstrating the versatility and utility of this technique even in non time dependent cases, and the accuracy of the perturbative order for our method.\footnote{We note, however, that we have restricted ourselves to odd metric perturbations, due to the difficulty in obtaining a master equation for even perturbations when matter is present.}

Whilst in this work we have only treated spherically symmetric background spacetimes and hence spherically symmetric ``dirt" around Schwarzschild black holes, our methods have the potential to be adapted for more complex scenarios. We are now extending our analysis to axisymmetric spacetimes, allowing us to consider perturbed Kerr spacetimes with axisymmetric matter clouds. This will allow us to treat more astrophysically relevant cases such as baryonic accretion disks. 

\section*{Acknowledgements}
\vspace{-0.2in}
\noindent We thank V Cardoso for helpful conversations. JB acknowledges funding from a UK Science and Technology Facilities Council (STFC) studentship. PF and KC acknowledge funding from the European Research Council (ERC) under the European Unions Horizon 2020 research and innovation programme (grant agreement No 693024). 

\appendix

\section{Perturbation Theory}
\label{App-perturbationtheory}

To find QNM for perturbed BH spacetimes we need two metric perturbations
\begin{equation}
    g_{\mu\nu} = g^{(0)}_{\mu\nu} + \epsilon g^{(1)}_{\mu\nu} + \zeta g^{(2)}_{\mu\nu},
\end{equation}
where $\epsilon$ and $\zeta$ are both small, but $\zeta \ll \epsilon$. The larger perturbation $\epsilon g^{(1)}_{\mu\nu}$ is static or slowly varying and captures the change to the metric from additional fields, modifications to GR, or the backreaction from clouds of matter. This perturbation is denoted $\delta g_{\mu\nu}$ earlier in previous sections. The smaller perturbation $\zeta g^{(2)}_{\mu\nu}$ is the one which will oscillate at the quasinormal mode frequencies. The $g^{(0)}_{\mu\nu}$ is a vaccuum background BH metric (Schwarzschild in this paper).

We have two sets of equations for the metric $g_{ab}$ and the matter fields $\mtf$: the Einstein field equations 
\begin{equation}
    G_{\mu\nu}[g_{ab}] = 8\pi T_{\mu\nu}[\mtf,g_{ab}],
\end{equation}
and the equation of motion
\begin{equation}
    \nabla_{\mu}[g_{ab}]T^{\mu\nu}[\mtf,g_{ab}] = 0.
\end{equation}
Here we will assume GR and assume $g^{(1)}_{\mu\nu}$ comes from the matter backreaction, and assume $T_{\mu\nu}$ to be order $\epsilon$ with $T_{\mu\nu} = \epsilon \Tilde{T}_{\mu\nu}$. We will also expand $\mtf$ as 
\begin{equation}
    \mtf = \mtf^{(0)} + \epsilon \mtf^{(1)} + \zeta \mtf^{(2)} +  \dots
\end{equation}
First let us expand in powers of $\zeta$. At order $\zeta^0$ we have
\begin{equation}
    G_{\mu\nu}[g^{(0)}_{ab}+\epsilon g^{(1)}_{ab}] = \epsilon 8\pi \Tilde{T}_{\mu\nu}[\mtf^{(0)}+\epsilon \mtf^{(1)},g^{(0)}_{ab}+\epsilon g^{(1)}_{ab}],
\end{equation}
and
\begin{equation}
    \epsilon\nabla_{\mu}[g^{(0)}_{ab}+\epsilon g^{(1)}_{ab}]T^{\mu\nu}[\mtf^{(0)}+\epsilon \mtf^{(1)},g^{(0)}_{ab}+\epsilon g^{(1)}_{ab}] = 0.
\end{equation}
We can then expand in powers of $\epsilon$. At order $\epsilon^0$ we have $G_{\mu\nu}[g^{(0)}_{ab}]=0$ as $g^{(0)}_{ab}$ is a vacuum solution. At order $\epsilon$ we have
\begin{equation}
    \frac{\delta G_{\mu\nu}}{\delta g_{ab}}[g^{(0)}_{pq}]g^{(1)}_{ab} = 8\pi \Tilde{T}_{\mu\nu}[\mtf^{(0)},g^{(0)}_{ab}], \label{zeta0_eq}
\end{equation}
and
\begin{equation}
    \nabla^{(0)}T^{\mu\nu}[\mtf^{(0)},g^{(0)}_{ab}] = 0,
\end{equation}
which can be solved for the zeroth order field solution $\mtf^{(0)}$ and the backreaction $g^{(1)}_{\mu\nu}$. At order $\zeta^1$ we obtain
\begin{align}
    &\frac{\delta G_{\mu\nu}}{\delta g_{ab}}[g^{(0)}_{pq}+\epsilon g^{(1)}_{pq}]g^{(2)}_{ab} = \nonumber \\
    &\qquad \epsilon 8\pi \frac{\delta \Tilde{T}_{\mu\nu}}{\delta\mtf}[\mtf^{(0)}+\epsilon \mtf^{(1)},g^{(0)}_{pq}+\epsilon g^{(1)}_{pq}]\mtf^{(2)} \nonumber \\
    &\qquad +\epsilon 8\pi \frac{\delta \Tilde{T}_{\mu\nu}}{\delta g_{ab}}[\mtf^{(0)}+\epsilon \mtf^{(1)},g^{(0)}_{pq}+\epsilon g^{(1)}_{pq}] g^{(2)}_{ab}+\dots, \nonumber \\
    &= \epsilon 8\pi \left(\frac{\delta \Tilde{T}_{\mu\nu}}{\delta\mtf}[\mtf^{(0)},g^{(0)}_{pq}]\mtf^{(2)}+\frac{\delta \Tilde{T}_{\mu\nu}}{\delta g_{ab}}[\mtf^{(0)},g^{(0)}_{pq}] g^{(2)}_{ab}\right) +\mathcal{O}(\epsilon^2).
\end{align}
Then let $\mtf^{(2)} = \Tilde{\varphi}_m^{(2)} + \chi^{(2)}$ such that 
\begin{equation}
    \frac{\delta \Tilde{T}_{\mu\nu}}{\delta\mtf}[\mtf^{(0)},g^{(0)}_{pq}]\chi^{(2)}+\frac{\delta \Tilde{T}_{\mu\nu}}{\delta g_{ab}}[\mtf^{(0)},g^{(0)}_{pq}] g^{(2)}_{ab} = 0.
\end{equation}
Then 
\begin{align}
    \frac{\delta G_{\mu\nu}}{\delta g_{ab}}[g^{(0)}_{pq}+\epsilon g^{(1)}_{pq}]g^{(2)}_{ab} =
    \frac{\delta \Tilde{T}_{\mu\nu}}{\delta\mtf}[\mtf^{(0)},g^{(0)}_{pq}]\Tilde{\varphi}_m^{(2)},
    \label{pert_eq} \\
    \epsilon\nabla_{\mu}[g^{(0)}_{pq}]\left(\frac{\delta T^{\mu\nu}}{\delta\mtf}[\mtf^{(0)},g^{(0)}_{ab}]\Tilde{\varphi}_m^{(2)}\right) = 0. \label{eom}
\end{align}
to order $\epsilon$. We see that \eqref{pert_eq} and \eqref{eom} can be written as
\begin{align}
    \mathcal{L}^{ab}_{\mu\nu} g^{(2)}_{ab} &= S_{\mu\nu} \Tilde{\varphi}_m^{(2)}, \label{pert_source} \\
    \mathcal{E}\Tilde{\varphi}_m^{(2)} &= 0. \label{varphi_perturb_eom}
\end{align}
where $\mathcal{E},\mathcal{L}^{ab}_{\mu\nu},S_{\mu\nu}$ are differential operators. Eq. \eqref{varphi_perturb_eom} provides an equation of motion for $\Tilde{\varphi}_m^{(2)}$, which then provides a source term $S_{\mu\nu}\Tilde{\varphi}_m^{(2)}$ to the equation of motion for the metric perturbation $g^{(2)}_{\mu\nu}$. The perturbed quasinormal modes are solutions of the homogeneous, unsourced equation 
\begin{equation}
    \mathcal{L}^{ab}_{\mu\nu} g^{(2)}_{ab} = 0.
\end{equation}
When decomposed into odd tensor harmonics, Eq. \eqref{pert_source} gives Eq. \eqref{Master_eq} where 
the source term $\Src$ is explicitly (see for example \cite{Martel_2005}) 
\begin{align}
    \Src^{lm} &= - r^2 \frac{l(l+1)}{(l-1)(l+2)} \epsilon^{ab} \Tilde{\nabla}_a \int \Delta T_{Ab} ~ (X^A_{lm})^* \dd \Omega, \\
    X^A_{lm} &= - \epsilon^{AB} \hat{\nabla}_B Y^{lm}(\theta, \phi), \\
    \Delta T_{\mu\nu} &= \frac{\delta T_{\mu\nu}}{\delta \mtf} \Tilde{\varphi}_m^{(2)}.
\end{align}
where $\epsilon^{ab}$ is the Levi-Civita symbol, $\hat{\nabla}_A$ is the covariant derivative on the 2-sphere, `` ${}^*$ " denotes complex conjugation and $Y^{lm}(\theta,\phi)$ are the complex spherical harmonics.

\section{Method in Schwarzschild coordinates}
\label{sec:SchMethod}

If the metric perturbation is time independent and sufficiently well behaved near the horizon we do not need to introduce ingoing EF coordinates and can instead repeat the derivation in the more familiar Schwarzschild coordinates, which we shall do now.

Let us again consider a perturbed Schwarzschild background metric of the form
\begin{equation}
ds^2=\;-(f+\delta f)\dd t^2+ (f+\delta g)^{-1} \dd r^2+r^2 \dd \Omega \label{modmetric},
\end{equation}
where we require $\delta f(r), \delta g(r) \ll 1$ in Schwarzschild coordinates. For odd modes in general, and for even modes in vacuum, this gives a modified QNM master equation of the form
\begin{align}
&\Big[F_\ast \partial_r ( F_\ast \partial_r)+(\omega^2-F ~V_\ast)\Big]\Psi=0\label{perteq1},\\
&F := f + \delta f, \\
& F_\ast(r) := \sqrt{(f+\delta f)(f+\delta g)} \approx \;1-\frac{2M}{r}+\frac{\delta f(r) + \delta g(r)}{2},\\
&V_\ast(r) = V_{\pm}(r) + \delta V(r).
\end{align}
where again $V_{\pm}$ corresponds to even/odd modes respectively and we have let $\Psi(t,r) = e^{-i\omega t}\Psi(r)$. In Eq. \eqref{perteq1} we allow the $\delta V$ to be an arbitrary function of $r$, containing both the terms arising from metric perturbations $\delta f, \delta g$, as we saw in the previous section, as well as from (for example) modified gravity effects. The only requirement we will impose is that $\delta V$ is small compared to the zeroth order potential $V_{\pm}$. If $\delta g \neq 0$ the location of the BH horizon will be shifted to
\begin{align}
r_H = 2M\left[1-\delta g(2M)\right].
\end{align}
We can then rewrite $F_\ast(r)$ as
\begin{align}
F_\ast(r) =&  f_H(r) Z(r), \\
f_H(r) :=& \left(1-\frac{r_H}{r}\right), \\
Z(r) =& 1 + \delta Z(r),\\
\delta Z(r) =& \; \frac{r\frac{\delta f(r) + \delta g(r)}{2}-2M\delta g(2M)}{r-2M},
\end{align}
again working to first order in all perturbed quantities. Note that for $\delta Z$ to be well behaved at $r=2M$ we need $\delta f(2M) = \delta g(2M)$. In the case with no modified gravity and with $g(2M) = f(2M) = 0$ we can directly compare to the expressions from section \ref{sec:perturbed_QNM_eq} and find
\begin{equation}
    \delta \lambda = (\delta f - \delta g)/(2f),
\end{equation}
and 
\begin{equation}
    \delta V = \frac{f}{r}\left(\frac{2\delta Z}{r} - \delta Z'\right)-\frac{2 \delta \lambda}{r^2}\left(1 - \frac{3M}{r}\right).
\end{equation}
If we again define $\Phi := \sqrt{Z}\Psi$ Eq. \eqref{perteq1} can be rewritten as 
\begin{equation}
	f_H \pdv{r}\left[ f_H \pdv{\Phi}{r}\right] + \left[\frac{\omega^2}{Z^2} - f_H V\right]\Phi = 0, 
\end{equation}
where to first order in $\delta$
\begin{equation}
	V = V_{\pm} + \delta V + (\delta f - 2 \delta Z) V_{\pm} + \tfrac{1}{2}(f_H \delta Z')'.
\end{equation}
and we seperate the $\omega$ term as 
\begin{equation}
\begin{split}
	\omega/Z^2 &= \omega^2[1 - 2\delta Z(r_H)] - 2\omega_0^2[\delta Z(r) - \delta Z(r_H)], \\
	&= \Omega^2  - 2\omega_0^2[\delta Z(r) - \delta Z(2M)], 
\end{split}
\end{equation}
to first order in small quantities. We can also rewrite the potentials $V_\pm$ in terms of $r_H$:
\begin{subequations}
\begin{align}
V_+ =&\; \tilde{V}_+ + \delta \tilde{V}_+\\
V_- =&\; \tilde{V}_- + \delta \tilde{V}_- 
\end{align}
\end{subequations}
where
\begin{subequations}
\begin{align}
\tilde{V}_+ = &\; \frac{(\ell+2)(\ell-1)}{3r^2}+\frac{r_H}{r^3}+\frac{2(\ell+2)^2(\ell-1)^2(\ell^2+\ell+1)}{3(3r_H+(\ell+2)(\ell-1)r)^2}\\
\delta \tilde{V}_+ = &\; \delta g(2M)\left(\frac{2M}{r^3}-\frac{8M(\ell+2)^2(\ell-1)^2(\ell^2+\ell+1)}{3(6M+(\ell+2)(\ell-1)r)^3}\right)\\
\tilde{V}_- = &\; \frac{\ell(\ell+1)}{r^2}+\frac{r_H}{r^3}(1-s^2)\\
\delta\tilde{V}_-=&\; \delta g(2M)\frac{2M}{r^3}(1-s^2).
\end{align}
\end{subequations}
Finally we obtain
\begin{align}
f_H\pdv{r}\left[f_H\pdv{\Phi}{r}\right]+\left[\Omega^2-f_H\left(\tilde{V}_{\pm}+\Delta V\right)\right]\Phi = 0\label{perteq2}
\end{align}
where $\Delta V$ again collects all the order $\delta$ terms, both from the original potential perturbation $\delta V$ as well as from modified geometry terms. Explicitly, 
\begin{equation}
\begin{split}
\Delta V =& \;\delta V + \delta \tilde{V}_\pm + \tilde{V}_\pm \left(\delta f/f_H - 2\delta Z\right)\\
& + \tfrac{1}{2}(f_H \delta Z')'+\frac{2\omega_0^2}{f_H}\left[\delta Z(r)-\delta Z(2M)\right].
\end{split}
\end{equation}
Unlike in the time dependent case $\omega$ and $\Omega$ are related by a simple constant rescaling
\begin{align}
\Omega = \omega\left(1-\delta Z(2M)\right).
\end{align}
We can relate $\delta Z(2M)$ to $\delta f, \delta g$ through use of l'H\^{o}pital's rule:
\begin{equation}
\delta Z(2M)=\delta g(2M)+ M\left[\delta f'(2M)+\delta g'(2M)\right].
\end{equation}

If we then again apply the method of \cite{Dolan_2009} we find solutions of the form
\begin{align}
\omega = \omega_0 + \omega_0 \delta Z(2M) + \delta \Omega (\Delta \ef).\label{omegafinal}
\end{align}
from which we can check that the perturbative expressions for $\delta \Omega(\Delta V)$ match those we derived in the main text with $\A$ set to zero. 

\section{Testing the method}
\label{sec:simple_examples} 
We do not have numerical results for the QNM perturbations of a growing dirty black hole to compare to the analytic results derived in the previous section. However we can apply the same techniques to several other examples in Schwarzschild coordinates for which results have been previously obtained - specifically, power law potentials, exponential potentials, Reissner-Nordstrom and de Sitter.
In particular, we confirm that the 5th order perturbative expansion of the frequency shift should be sufficiently accurate, and that higher corrections will not significantly change the result.

\subsection{Power Law Potentials}
\begin{table*}
\caption{Comparison between the analytic results presented here and the numeric results of Cardoso et al for the $\ell=2$ odd parity gravitational QNM deviation $\Delta\omega$ to order $L^{-8}$.}
\label{integertablel2gravodd}
\begin{ruledtabular}
\begin{tabular}{ccccc}
 $p$ & $2M\Delta\omega$ (analytic)  & $2M\Delta\omega$ (numeric)  & \% error $\Delta\omega_R$ & \% error $\Delta\omega_I$
 \\
\hline
 0 & 0.243747+0.0913876$i$ & 0.247252+0.0926431$i$ & -1.41747 & -1.3552 \\
 1 & 0.158967+0.0180090$i$ & 0.159855+0.0182085$i$ & -0.555267 & -1.09556 \\
 2 & 0.0966513-0.00277561$i$ & 0.0966322-0.0024155$i$ & 0.019719& 14.9086 \\
 3 & 0.0585225-0.00410688$i$ & 0.0584908-0.00371786$i$ & 0.0542632& 10.4635 \\
 4 & 0.0366465-0.000745599$i$ & 0.0366794-0.000438698$i$ & -0.0896896& 69.9573 \\
 5 & 0.0240123+0.00249465$i$ & 0.0240379+0.00273079$i$ & -0.106785& -8.64721 \\
\end{tabular}
\end{ruledtabular}
\end{table*}

\begin{table*}
\caption{Comparison between the analytic results presented here and the numeric results of Cardoso et al for the $\ell=2$ even parity gravitational QNM deviation $\Delta\omega$ to order $L^{-8}$.}
\label{integertablel2graveven}
\begin{ruledtabular}
\begin{tabular}{ccccc}
 $p$ & $2M\Delta\omega$ (analytic)  & $2M\Delta\omega$ (numeric)  & \% error $\Delta\omega_R$ & \% error $\Delta\omega_I$
 \\
\hline
 0 & 0.224732+0.0916972 $i$ & 0.22325+0.09312 $i$ & 0.663953 & -1.52787 \\
 1 & 0.153719+0.0195864 $i$ & 0.154195+0.019927$i$i & -0.308834 & -1.70931 \\
 2 & 0.0974921-0.00328011 $i$ & 0.0978817-0.0034275 $i$ & -0.398015 & -4.30028 \\
 3 & 0.0614226-0.00618217 $i$ & 0.0616142-0.0064403 $i$ & -0.310969 & -4.00799 \\
 4 & 0.0399055-0.00334236 $i$ & 0.0400156-0.0036191 $i$ & -0.275227 & -7.64665 \\
 5 & 0.0271051+0.0000307656 $i$ & 0.0271849-0.0002403 $i$ & -0.293483 & -112.803 \\
\end{tabular}
\end{ruledtabular}
\end{table*}
\begin{table*}[t]
\caption{Comparison between the analytic results presented here and the numeric results of Cardoso et al. (2019) \cite{Cardoso_2019} for an exponential potential deviation for the even parity gravitational modes to order $L^{-8}$.}
\label{exptablegraveven}
\begin{ruledtabular}
\begin{tabular}{ccccc}
 $\ell$ & $2M\Delta\omega$ (analytic)  & $2M\Delta\omega$ (numeric)  & \% error $\Delta\omega_R$ & \% error $\Delta\omega_I$
 \\
\hline
 2 & 0.439353+0.108479 $i$ & 0.438579+0.110111 $i$ & 0.176484 & -1.48256 \\
 3 & 0.274923+0.0447816 $i$ & 0.274902+0.0448262 $i$ & 0.00780032 & -0.0996234 \\
 4 & 0.202828+0.0250234 $i$ & 0.202826+0.0250268 $i$ & 0.000874023 & -0.013268 \\
 5 & 0.161632+0.0161213 $i$ & 0.161632+0.0161217 $i$ & 0.000123541 & -0.002308 \\
\end{tabular}
\end{ruledtabular}
\end{table*}
First we will try potential deviations $\delta V$ about a pure Schwarzschild background, such that $\delta f = \delta g = 0$. To compare with the numerical results of Cardoso et al. (2019) \cite{Cardoso_2019}, we will first assume the following form for the potential deviations (for both Regge-Wheeler and Zerilli type equations):
\begin{align}
\delta V = \frac{\alpha}{(2M)^2} \left(\frac{2M}{r}\right)^p,\; p \geq 0. \label{powerlawpot}
\end{align}
and $\alpha$ is a dimensionless constant. In Cardoso et al., $p$ is assumed to be an integer and numeric results for QNM deviations are provided for values of $p$ from 0 to 50. In this section we will first compare the analytic results at integer values to the Cardoso et al. values before allowing $p$ to vary continuously.  

Tables \ref{integertablel2gravodd} and \ref{integertablel2graveven} show a comparison for the first few values of $p$ for the odd and even parity $\ell=2$ gravitational QNMs respectively to order $L^{-8}$. We see that very good agreement between the two methods is found in the real part of the frequency deviation $\Delta\omega_R$, with slightly worse agreement in the imaginary part $\Delta\omega_I$. Note that some of the percentage errors can be misleading when the values of the deviations are extremely close to 0, for example in the case of the $p=5$ deviation for the $\ell=2$ even parity QNM.

We've seen that for potential deviations of the form given in Eq.~(\ref{powerlawpot}) the analytic QNM deviations presented here compare well with those calculated numerically as long as the index $p$ doesn't exceed around 15, though this `guide' is dependent on the angular harmonic index $\ell$ (good agreement is found for larger $p$ with high $\ell$) and on whether the perturbations are of scalar, vector, or gravitational type. 

\begin{figure*}[t]
\includegraphics[width=0.45\textwidth]{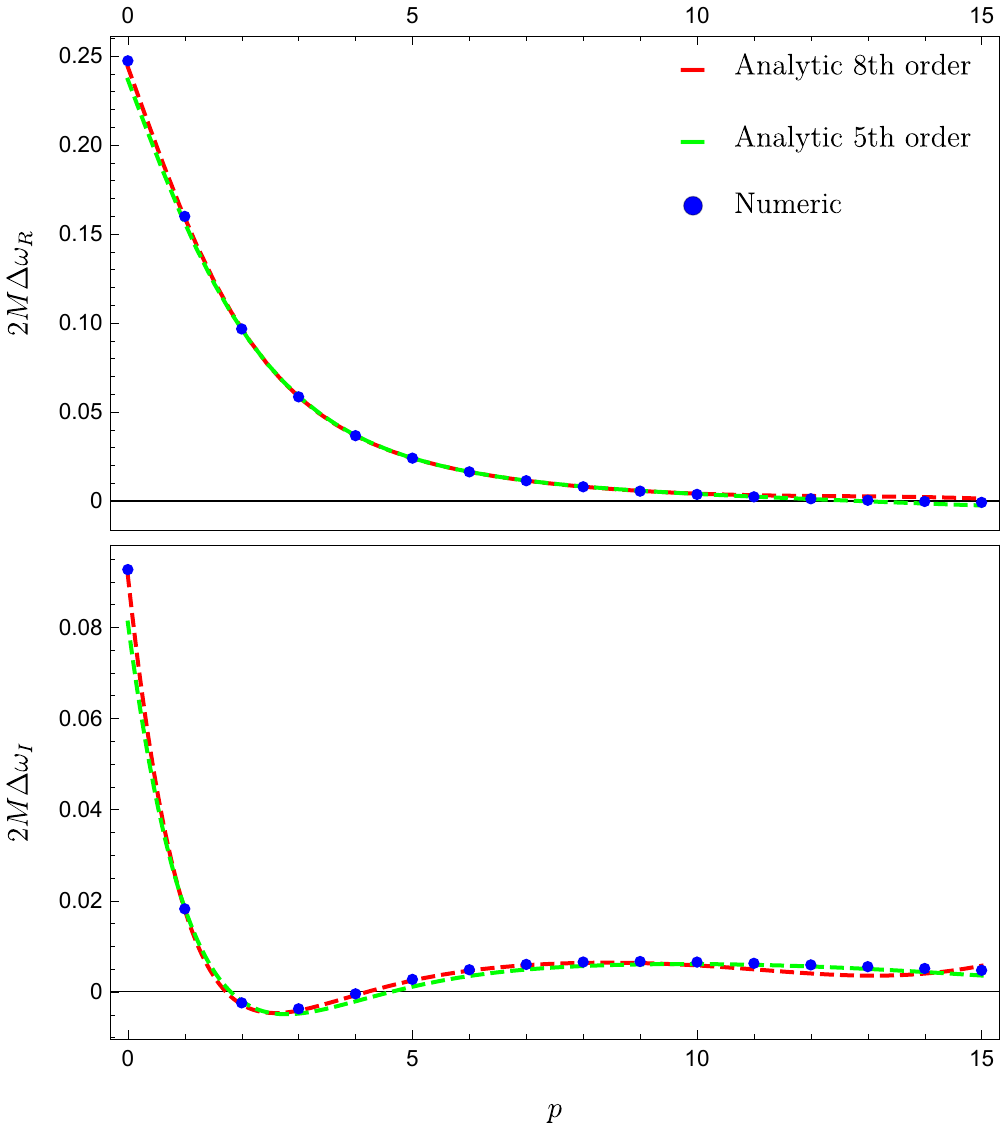}
\includegraphics[width=0.45\textwidth]{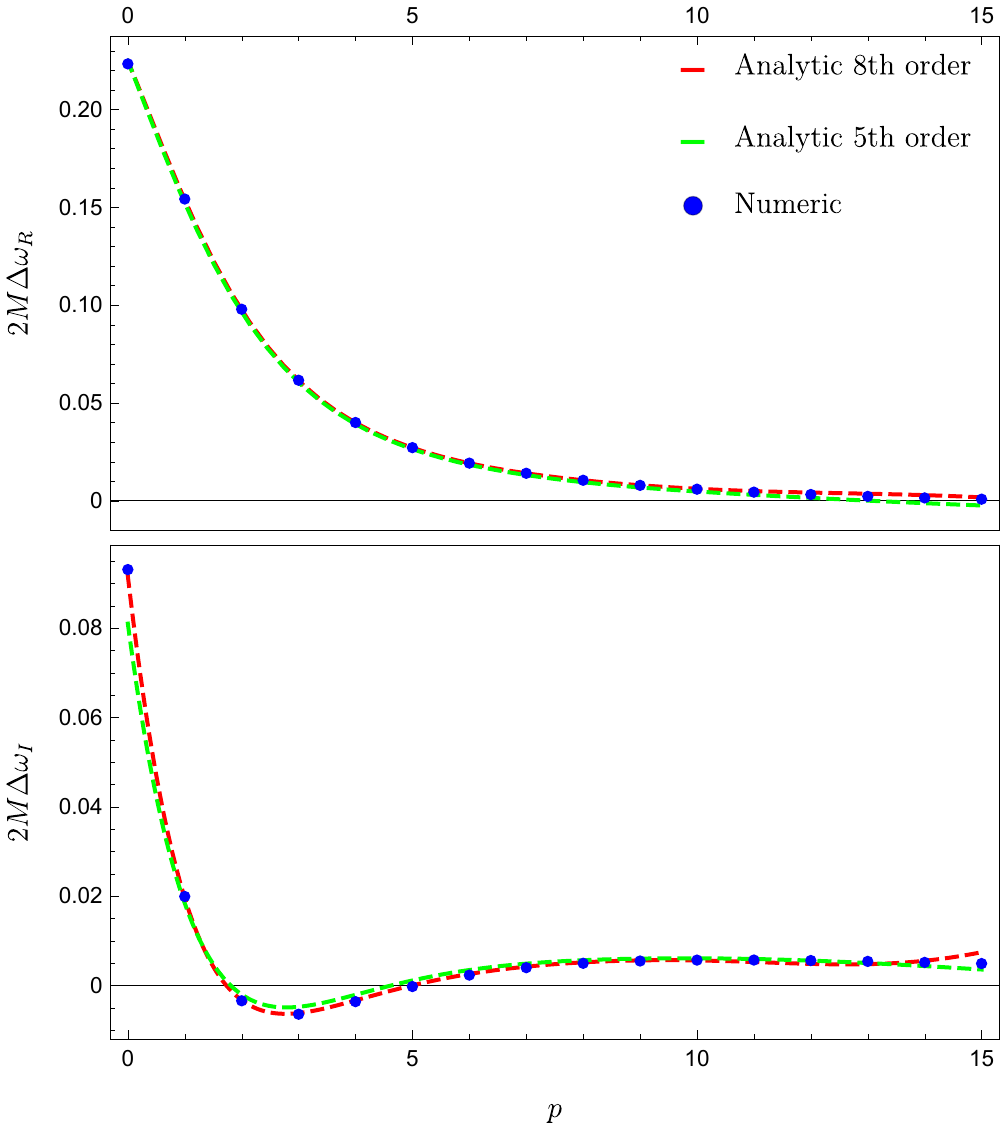}
\caption{Graphical comparison between the analytic results presented here and the numeric results of Cardoso et al.(2019) \cite{Cardoso_2019} for the $\ell=2$ odd (left panel) and even (right panel) parity $\Delta\omega$ to order $L^{-8}$ (red) and order $L^{-5}$ (green).}\label{integerfigl2gravodd}
\end{figure*}

Fig. \ref{integerfigl2gravodd} shows a plot of the numeric results of Cardoso et al with the analytic QNM deviations presented here, for both the $\ell=2$ odd and even parity gravitational QNMs. In this case we are allowing $p$ to be continuous for the analytic results. Good agreement is shown between the two methods up to around $p=10$, at which point the imaginary component of the analytic $\Delta\omega$ starts to visibly deviate from the numeric results. We find that above around $p=15$ large deviations from the numeric results are seen in both the $\Delta\omega_R$ and $\Delta\omega_I$, with the analytic curve showing oscillatory behaviour. We do not have an explanation for this shortcoming at the moment, so clearly the analytic results are best restricted for use up to $p \approx 15$. Similar behaviour is seen for vector and scalar perturbations, with better agreement between the analytic and numeric results found for larger values of $\ell$.

\subsection{Exponential Potential}

We will now study more unconventional potential deviations and again compare the analytic QNM deviation results with those calculated numerically. First, we consider the addition of an exponential function to the potential:
\begin{align}
\delta V = \frac{1}{(2M)^2}\exp \left(\frac{2M}{r}\right). \label{exponentialpot}
\end{align}
If we use the Taylor series representation of the exponential function we can write Eq.~(\ref{exponentialpot}) as a sum of integer powers of $2M/r$, thus allowing us to use the results of Cardoso et al. as a comparison for the analytic results. Table \ref{exptablegraveven} gives the deviations calculated with both methods for the $\ell = 2 - 5$ even parity gravitational modes, with extremely good agreement found between the two methods. 

\subsection{Reissner-Nordstr{\"o}m background}

Following Dolan \& Ottewill (2009) \cite{Dolan_2009} we can write the unperturbed master equation for the charged Reissner-Nordstr{\"o}m black hole as 
\begin{equation}
	f_q\pdv{r}\left[f_q\pdv{\Psi}{r}\right] + \left[\omega^2 - f_q V_{\pm}\right]\Psi = 0,
\end{equation}
where $q = Q/M$ the charge-to-mass ratio, $f_q(r) = 1 - 2M/r + q^2 M^2/r^2$, and the odd mode potential is 
\begin{equation}
	V_- = \frac{L^2 - 1/4}{r^2} - \frac{M\kappa_s}{r^3} + \frac{q^2 M^2\eta_s}{r^4},
\end{equation}
where 
\begin{equation}
	\eta_s, \kappa_s = \begin{cases}
	2, \; -2 & s = 0 \\
	4,\; 3-\sqrt{9+4 q^2 (L^2-9/4)} & s = 1 \\
	4,\; 3+\sqrt{9+4 q^2 (L^2-9/4)} & s = 2.
	\end{cases}
\end{equation}
The Reissner-Nordstr{\"o}m metric is 
\begin{equation}
	\dd s^2 = - f_q \dd t^2 + f_q^{-1} \dd r^2 + r^2 \dd \Omega.
\end{equation}
Consider the weakly charged case where $q \ll 1$. In that limit we can assign 
\begin{align}
\delta f(r) &= \delta g(r) = q^2M^2/r^2, \\  
\delta V(r) &= q^2\left[\frac{M(\tfrac{3}{2} - \tfrac{2}{3}L^2)}{r^3} + \frac{4M^2}{r^4}\right].
\end{align}
This gives for the $n=0$, $l=2$ odd mode (again to order $L^{-8}$)
\begin{equation}
	\omega_{QNM} =\omega_0 + \frac{(0.0252499 - 0.00267011 i) q^2}{M} + \mathcal{O}(q^3), 
\end{equation}
which compares favourably to the numerical result from Cardoso et al. of 
\begin{equation}
	\omega_{QNM} =\omega_0 + \frac{(0.0258177 - 0.002824 i) q^2}{M} + \mathcal{O}(q^3), 
\end{equation}
with a relative difference of $2.2\%, 5.4\%$ between the two for the for the real and imaginary parts respectively. 

\subsection{de Sitter background}

In Schwarzschild de Sitter (SdS) spacetime the line element takes the form 
\begin{equation}
	\dd s^2 = - f_{\delta \Lambda} \dd t^2 + f_{\delta \Lambda}^{-1} \dd r^2 + r^2 \dd \Omega
\end{equation}
where $f_{\delta \Lambda}(r) = 1 - 2M/r - \delta \Lambda r^2/3$. The master equation is 
\begin{equation}
	f_{\delta \Lambda}\pdv{r}\left[f_{\delta \Lambda}\pdv{\Psi}{r}\right] + \left[\omega^2 - f_{\delta \Lambda} V_{\pm}\right]\Psi = 0,
\end{equation}
where $V_{\pm}$ are the standard Zerilli and Regge-Wheeler potentials. Let $\delta \Tilde{\Lambda} = \delta \Lambda M^2$. If we take $\delta  \Tilde{\Lambda} \ll 1$ we can proceed as before 
\begin{equation}
	\delta f(r) = \delta g(r) = - \delta \Lambda r^2 /3, \quad \delta V(r) = 0. 
\end{equation}
For the $n=0$, $l=2$ odd mode to order $L^{-8}$ we find
\begin{equation}
	\omega_{QNM} = \omega_0 + (-1.67328 + 0.332735 i)\delta \Tilde{\Lambda}/M + \mathcal{O}({\delta \Tilde{\Lambda}}^2).
\end{equation}
The equivalent calculation for non-linear $\delta \Lambda$ dependence to order $L^{-6}$ (see Tattersall (2018) \cite{Tattersall_2018})) gives 
\begin{equation}
\begin{split}
	\omega_{QNM} =& \omega_0 + (-1.67328 + 0.332735 i) \delta \Tilde{\Lambda} /M + \\
	&(-3.90506 +1.15466 i) {\delta \Tilde{\Lambda}}^2/M + \\
	&(-16.9027 +5.13741 i) {\delta \Tilde{\Lambda}}^3 /M + \mathcal{O}({\delta \Tilde{\Lambda}}
^4),
\end{split}
\end{equation} 
so for small $\delta \Tilde{\Lambda}$ we have excellent agreement with the non-linear calculation with a fraction of the effort. Note that this is mathematically equivalent to the case of non-accreting uniform density dark matter as described in \cite{Barausse:2014tra} equation (67) with $\delta \Lambda = 8 \pi \rho_{DM}$.

\bibliography{bibliography}

\end{document}